\documentclass[12pt]{article}
\usepackage{amsmath, amssymb, amsthm}
\usepackage{xcolor}
\usepackage{verbatim}
\usepackage{authblk}
\usepackage{natbib}
\usepackage{tikz}

\title{\bf A Note on the Identifiability of the Degree-Corrected Stochastic Block Model}
 \author[1]{John Park}
  \author[2]{Yunpeng Zhao}
  \author[1,3,*]{Ning Hao}
  \affil[1]{Department of Mathematics, the University of Arizona}
  \affil[2]{Department of Statistics, Colorado State University}
  \affil[3]{Statistics \& Data Science GIDP, the University of Arizona}
  \affil[*]{Corresponding author: nhao@arizona.edu}
\date{}

\begin{document}
\newtheorem{theorem}{Theorem}
\newtheorem{lemma}{Lemma}
\newtheorem{proposition}{Proposition}
\newtheorem{condition}{Condition}
\newtheorem{example}{Example}
\newcommand{\diag}{\mathrm{diag}}
\newcommand{\rank}{\mathrm{rank}}

\newcommand{\tcb}[1]{\textcolor{blue}{#1}}
\newcommand{\tcr}[1]{\textcolor{red}{#1}}

\maketitle

\begin{abstract}
In this short note, we address the identifiability issues inherent in the Degree-Corrected Stochastic Block Model (DCSBM). We provide a rigorous proof demonstrating that the parameters of the DCSBM are identifiable up to a scaling factor and a permutation of the community labels, under a mild condition.  
\end{abstract}
\noindent%

\section{Introduction}
The Stochastic Block Model (SBM) is a foundational framework for modeling community structures in networks. It groups nodes into distinct communities, with the probability of an edge between two nodes determined solely by their community memberships \citep{HolLei81,Nowicki2001,Bickel&Chen2009,choi2012stochastic}. SBM extends the Erd\H{o}s-R\'{e}nyi model \citep{erdds1959random}, which assumes a uniform edge probability across all node pairs, thereby offering a simpler framework without community differentiation. In contrast, SBM captures both inter-community and intra-community connection patterns, making it a valuable tool for community detection \citep{abbe2017community,zhao2017survey}. However, SBM assumes homogeneity within each community, limiting its applicability to real-world networks where significant degree variability exists among nodes within the same community. To address this limitation, the Degree-Corrected Stochastic Block Model (DCSBM) was introduced. DCSBM extends SBM by incorporating degree heterogeneity, allowing nodes within a community to exhibit varying degrees of connectivity \citep{karrer2011stochastic,ball2011efficient}. This enhancement improves the model’s flexibility, making DCSBM particularly suited for networks such as social, biological, and communication systems, where nodes often exhibit varying connectivity patterns even within the same community.

It is well known that parameters in an SBM, like those in other clustering models such as Gaussian mixture models, are not identifiable due to the possible permutation of community labels.  For the DCSBM, this issue is further compounded by the non-identifiability of degree parameters due to scaling factors \citep{karrer2011stochastic,ball2011efficient, zhao2012consistency}.
Despite these challenges, researchers generally agree that these identifiability issues pose only minor technical obstacles and that no additional identifiability problems exist for the DCSBM. However, formal references addressing the identifiability of the DCSBM are limited.  In this note, we point out a small gap in a commonly used argument and provide a rigorous proof confirming that the parameter system of the DCSBM is identifiable up to a scaling factor and a permutation of the community labels, under a mild condition that each community has at least three members. To our best knowledge, this condition has never been explicitly stated in the literature.  In addition, we show that this condition is necessary using a counterexample. Together, our results fill a critical gap in the theoretical understanding of the identifiability of the DCSBM.

\section{Model Description}
Consider an undirected graph \( G = (V, E) \) with \( n \) nodes. The adjacency matrix \( A \) represents the presence or absence of edges between nodes. The DCSBM with \(n\) nodes and \(K\) communities is characterized by the following components with technical conditions.

\begin{enumerate}
    \item \textbf{Community Membership Matrix \( Z \)}:
    \( Z \) is an \( n \times K \) binary matrix where \( Z_{ik} = 1 \) if node \( i \) belongs to community \( k \), and 0 otherwise. We require that each community has at least 1 member. We use \(z_i\) to denote the community label of node \(i\). That is, \(z_i=k\) if and only if \(Z_{ik}=1\). 

    \item \textbf{Degree Parameter Matrix \( \Theta \)}:
    \( \Theta \) is an \( n \times n \) diagonal matrix where \( \Theta_{ii} = \theta_i>0 \) represents the degree parameter for node \( i \).

    \item \textbf{Community Connection Matrix \( B \)}:
    \( B \) is a symmetric full rank \( K \times K \) matrix where \( B_{kl} \) represents the intensity of connection between nodes in community \( k \) and community \( l \).
\end{enumerate}

A triple \(\{Z,\Theta,B\}\) defines a DCSBM,  where the off-diagonal elements of the adjacency matrix \( A \) are generated independently with \(\mathbb{E}[A_{ij}]=\Delta_{ij} \), and 
\begin{equation}\label{formula1}
     \Delta = \Theta Z B Z^\top \Theta.
\end{equation}
We assume that all triples \(\{Z,\Theta,B\}\) satisfy the conditions listed above throughout this paper. 

Now, we introduce useful notations and technical lemmas. We use $[n]$ to denote the set of integers $\{1,\dots,n\}$. A partition of $[n]$ is a collection of nonempty, mutually disjoint subsets of $[n]$ whose union is $[n]$. A permutation matrix is a square binary matrix with exactly one entry of 1 in each row and each column, with all other entries being 0.  A $K \times K$ permutation matrix $P$ corresponds to a permutation of a set with $K$ elements. 
The operator \(\diag(\cdot)\) is defined as \(\diag: \mathbb{R}^{n}\to\mathbb{R}^{n\times n}\), mapping a vector to a diagonal matrix such that \([\diag(v)]_{ii}=v_i\) for a vector $v=(v_1,\dots,v_n)^{\top}$. Finally, we use $1_n$ to denote an  $n$-dimensional vector where all entries are equal to 1. 

Every equivalence relation on a set defines a partition of that set, and vice versa. For a matrix $Q$, we define its $i$-th row and $j$-th row to be equivalent if they are identical. Using this equivalence relation, any matrix $Q$ with $n$ rows defines a partition of $[n]$. The following lemmas will be useful in our theoretical derivations.

\begin{lemma}\label{lem1}
    An $n\times K$ community membership matrix $Z$ defines a partition of $[n]$ based on row equivalence. Two community membership matrices $Z$ and $\tilde Z$ define the same partition if and only if $\tilde Z=ZP$ for a permutation matrix $P$. 
\end{lemma}

\noindent \textbf{Proof of Lemma \ref{lem1}:} Define subset $C_k\subset [n]$ as $\{i\in[n]\mid Z_{ik}=1\}$. It is straightforward to verify that $\{C_1,\dots, C_K\}$ is a partition of $[n]$. Similarly, we can define another partition $\{\tilde C_1,\dots,\tilde C_K\}$ based on $\tilde Z$. The partitions $\{C_1,\dots, C_K\}$ and $\{\tilde C_1,\dots,\tilde C_K\}$ are identical if and only if there exists a permutation $\pi:[K]\to[K]$ such that $\tilde C_{\pi(k)}=C_k$.  This permutation $\pi$ induces a permutation matrix $P$ with $P_{k,\pi(k)}=1$,  and it follows that $\tilde Z=ZP$.  \hfill$\Box$

\begin{lemma}\label{lem2}
    Let $Z$ and $\tilde Z$ be two $n\times K$ community membership matrices, and $R$ and $\tilde R$ be two full row-rank $K\times K'$ matrices where $K'\geq K$. The matrices $ZR$ and $\tilde Z\tilde R$ define the same partition of $[n]$ based on row equivalence if and only if $\tilde Z=ZP$ for a permutation matrix $P$. 
\end{lemma}
\noindent \textbf{Proof of Lemma \ref{lem2}:} All rows of $R$ are distinct, since $R$ has full row rank. The rows of $ZR$ are replications of rows of $R$, and the $i$-th row of $ZR$ is identical to the $k$-th row of $R$ if and only if $Z_{ik}=1$. As a result, $ZR$ and $Z$ define the same partition of $[n]$ based on row equivalence. The same conclusion holds for $\tilde{Z}\tilde{R}$ and $\tilde{Z}$. 

Thus, $ZR$ and $\tilde Z\tilde R$ define the same partition of $[n]$ if and only if $Z$ and $\tilde Z$ define the same partition, and by Lemma \ref{lem1}, if and only if $\tilde Z=ZP$ for a permutation matrix $P$. \hfill$\Box$

For a matrix $Q$, we define its $i$-th row and $j$-th row to be proportionally equivalent if they are proportional to each other up to a positive scaling factor. Any matrix $Q$ with $n$ rows defines a partition of $[n]$ via this proportional equivalence.  

\begin{lemma}\label{lem3}
    The matrices $Z$, $\tilde Z$, $R$ and $\tilde R$ are defined as in Lemma \ref{lem2}. $\Theta$ and $\tilde{\Theta}$ are two $n\times n$ positive diagonal matrices. The matrices $\Theta ZR$ and $\tilde\Theta \tilde Z\tilde R$ define the same partition of $[n]$ based on row proportional equivalence if and only if $\tilde Z=ZP$ for a permutation matrix $P$. Moreover, if $\Theta ZR=\tilde\Theta \tilde Z\tilde R$, and $\{C_1,\dots,C_K\}$ is the partition defined by $\Theta ZR$ and $\tilde\Theta \tilde Z\tilde R$, then $\Theta_{ii}^{-1}\tilde{\Theta}_{ii}=\Theta_{jj}^{-1}\tilde{\Theta}_{jj}$ whenever $i,j\in C_k$.
\end{lemma}
\noindent \textbf{Proof of Lemma \ref{lem3}:} No two rows of $R$ are proportional to each other since $R$ has full row rank. Therefore, the partition defined by $ZR$ via row equivalence and the partition defined by $\Theta ZR$ via row proportional equivalence are the same, and the first conclusion follows Lemma \ref{lem2}.  To show the second statement, we write    \[ZR=\Theta^{-1}\tilde\Theta \tilde Z\tilde R, \]
where $\Theta^{-1}\tilde\Theta$ is a positive diagonal matrix. It is straightforward to check when $i,j\in C_k$, the $i$-th and $j$-th rows of $ZR$ are identical, and the $i$-th and $j$-th rows of $\tilde Z\tilde R$ are identical, which implies $\Theta_{ii}^{-1}\tilde{\Theta}_{ii}=\Theta_{jj}^{-1}\tilde{\Theta}_{jj}$. \hfill $\Box$

\section{The Identifiability Issue on the DCSBM}
\subsection{A folklore}
In the literature on model identifiability, the expected values of the diagonal entries of $A$ are typically assumed to follow the same form as the off-diagonal entries, i.e., 
$\mathbb{E}[A] =  \Delta =\Theta Z B Z^\top \Theta$. Under this assumption, the derivations in \cite{karrer2011stochastic} suggest that the parameters are identifiable up to $K$ degrees of freedom, corresponding to $K$ scaling factors. The following proposition, though rarely stated explicitly in the literature, has been employed to provide a resolution to the identifiability issue of the DCSBM. 

\begin{proposition}\label{prop1}
Two parameter systems \(\{Z, \Theta, B\}\) and \(\{\tilde{Z}, \tilde{\Theta}, \tilde{B}\}\) satisfy 
\begin{equation}\label{formula2}
\Theta Z B Z^\top \Theta = \tilde{\Theta} \tilde{Z} \tilde{B} \tilde{Z}^\top \tilde{\Theta} =\Delta.
\end{equation}
if and only if \(\tilde{Z}=ZP\), \(\tilde{B}=P^{\top}DBDP\), and \(\tilde{\Theta}=\Theta \diag\left(ZD^{-1}1_K\right)\) where \(P\) is a \(K\times K\) permutation matrix, and \(D\) is a \(K\times K\) positive diagonal matrix.
\end{proposition}

In this proposition, \(\tilde{Z}=ZP\) indicates the membership matrices in two parametrization systems are up to a permutation \(P\). When two systems use the same labels, i.e., \(P=I_K\), they represent the same DCSBM if and only if \(\tilde{B}= DBD\), and \(\tilde{\Theta}=\Theta \diag\left(ZD^{-1}1_K\right)\) for a diagonal matrix \(D\) containing $K$ scaling factors, one for each community.

\vspace{0.5cm}

\noindent \textbf{Proof of Proposition \ref{prop1}:} To show the ``if'' part, 
we calculate
\begin{align*}
\tilde{\Theta} \tilde{Z} \tilde{B} \tilde{Z}^\top \tilde{\Theta} 
&= \Theta \diag\left(ZD^{-1}1_K\right) ZPP^{\top}DBDPP^{\top}Z^{\top}\diag\left(ZD^{-1}1_K\right)\Theta\\
&=\Theta \diag\left(ZD^{-1}1_K\right) ZDBDZ^{\top}\diag\left(ZD^{-1}1_K\right)\Theta.
\end{align*}
To obtain equation \eqref{formula2}, it suffices to show \[\diag\left(ZD^{-1}1_K\right) ZD= Z.\] We verify it by checking each entry. As both $\diag\left(ZD^{-1}1_K\right)$ and $D$ are  diagonal matrices, the \((i,k)\) entry on the left is  
\[ [\diag\left(ZD^{-1}1_K\right) ZD]_{ik} = [\diag\left(ZD^{-1}1_K\right)]_{ii} Z_{ik}D_{kk}. \]
When $z_i\ne k$, both \([\diag\left(ZD^{-1}1_K\right) ZD]_{ik}\) and \(Z_{ik}\) are zero. When $z_i= k$, \(Z_{ik}=1\), it is easy to check \([\diag\left(ZD^{-1}1_K\right)]_{ii}=D_{kk}^{-1}\), so \[[\diag\left(ZD^{-1}1_K\right)]_{ii} Z_{ik}D_{kk} = D_{kk}^{-1}Z_{ik}D_{kk} =Z_{ik}.\]

Now we show the ``only if'' part. 
First, we calculate the rank of \(\Delta = \Theta Z B Z^\top \Theta\) as
\[\rank (\Delta)=\rank (ZBZ^{\top})=\rank (B)=K.\]
The first equal sign follows that $\Theta$ is a positive definite diagonal matrix. The second equal sign holds because the membership matrix \(Z\) contains \(K\times K\) identity matrix \(I_K\) as a submatrix (up to a permutation).

Now consider the eigenvalue decomposition \(\Delta=U\Lambda U^{\top}\), where \(\Lambda\) is a \(K\times K\) diagonal matrix that contains all nonzero eigenvalues of $\Delta$, \(U\) is a $n\times K$ matrix whose columns are eigenvectors corresponding to nonzero eigenvalues. By the eigenvalue decomposition, we can write
\begin{equation}\label{formula3}
U=\Delta U \Lambda^{-1}=\Theta Z B Z^\top \Theta U \Lambda^{-1}=\Theta Z R,    
\end{equation}
where \(R=B Z^\top \Theta U \Lambda^{-1}\) is a full rank $K\times K$ matrix.  We conduct similar operations with $\{\tilde Z,\tilde\Theta,\tilde B\}$, and express $U$ via both parameter system 
\begin{equation}\label{formula4}
 U= \Theta Z R= \tilde \Theta\tilde Z \tilde R,
\end{equation}
where \(\tilde R=\tilde B \tilde Z^\top \tilde \Theta U \Lambda^{-1}\). \eqref{formula4} implies that $ \Theta Z R$ and $ \tilde \Theta\tilde Z \tilde R$ define the same partition via row proportional equivalence. By Lemma $\ref{lem3}$, we have $\tilde Z=ZP$ for a permutation matrix $P$. In addition, we denote the partition by $\{C_1,\dots,C_K\}$ where $C_k=\{i\in [n]\mid z_i=k\}$. Lemma $\ref{lem3}$ also implies that $\Theta_{ii}^{-1}\tilde{\Theta}_{ii}=\Theta_{jj}^{-1}\tilde{\Theta}_{jj}$, when $i,j\in C_k$.
 
Let \(D\) be a \(K\times K\) diagonal matrix, where \(D_{kk}=\tilde{\Theta}_{ii}^{-1}\Theta_{ii}\) for any $i$ with $z_i=k$, which is well-defined by the previous paragraph. It can be verified directly that $\Theta^{-1}\tilde{\Theta}=\diag\left(ZD^{-1}1_K\right)$, and $\diag\left(ZD^{-1}1_K\right)Z=ZD^{-1}$.  

Note that we have shown $\tilde Z=ZP$.  Equation \eqref{formula2} with $\tilde Z=ZP$ leads to
\begin{align*}
    ZBZ^\top&=\Theta^{-1}\tilde{\Theta}\tilde Z \tilde{B} \tilde Z^{\top}  \tilde{\Theta}\Theta^{-1}\\
    &=\Theta^{-1}\tilde{\Theta} ZP \tilde{B} P^{\top} Z^{\top} \tilde{\Theta}\Theta^{-1}\\
     &=\diag\left(ZD^{-1}1_K\right) ZP \tilde{B} P^\top Z^{\top} \diag\left(ZD^{-1}1_K\right)\\
     &=ZD^{-1}P \tilde{B} P^{\top} D^{-1}Z^{\top},
\end{align*}
which indicates \(B=D^{-1}P\tilde{B}P^{\top} D^{-1}\), or equivalently 
\(\tilde{B}=P^{\top}DBDP\). We have thus shown that if two parameter sets produce the same expected adjacency matrices, then \(\tilde{Z}=ZP\), \(\tilde{B}=P^{\top}DBDP\), and \(\tilde{\Theta}=\Theta \diag\left(ZD^{-1}1_K\right)\).
\hfill $\Box$

Proposition \ref{prop1} states that the equation \eqref{formula2} implies the community structure of the DCSBM is well-defined, only up to a label permutation. However, this result is insufficient to address the identifiability issue of the DCSBM,  where the diagonal elements in $\Delta$ are always assumed to be zero.

Let $\mathbb{P}$ denote the natural projection from the linear space of $n\times n$ matrices to the subspace of all $n\times n$ matrices with zero diagonals. In the DCSBM, it is typically assumed that, instead if \eqref{formula1}, the expected adjacency matrix satisfies
\begin{equation*}
     \mathbb{E}[A] = \mathbb{P} (\Theta Z B Z^\top \Theta).
\end{equation*}
In other words, two systems \(\{Z, \Theta, B\}\) and \(\{\tilde{Z}, \tilde{\Theta}, \tilde{B}\}\) define the same DCSBM if and only if
\begin{equation}\label{formula5}
\mathbb{P} (\Theta Z B Z^\top \Theta) = \mathbb{P} (\tilde{\Theta} \tilde{Z} \tilde{B} \tilde{Z}^\top \tilde{\Theta}).
\end{equation}
It is important to note that equation \eqref{formula5} imposes a weaker constraint than \eqref{formula2}. Consequently, Proposition \ref{prop1} does not suffice to fully resolve the identifiability issue in the DCSBM. The counterexamples provided below further illustrate this point.

\subsection{Counterexamples}
We illustrate two examples, which show that equation \eqref{formula5} itself cannot enforce the identifiability of the DCSBM. 
\begin{example}
Consider two parameter sets with different community structures:
\begin{align*}
\Theta=\begin{bmatrix}2&0&0\\0&2&0\\0&0&2\end{bmatrix},\qquad Z&=\begin{bmatrix}1&0\\0&1\\0&1\end{bmatrix},\qquad B=\frac{1}{40}\begin{bmatrix}2&1\\1&2\end{bmatrix};\\
\tilde{\Theta}=\begin{bmatrix}1&0&0\\0&2&0\\0&0&4\end{bmatrix},\qquad \tilde{Z}&=\begin{bmatrix}1&0\\1&0\\0&1\end{bmatrix},\qquad \tilde{B}=\frac{1}{40}\begin{bmatrix}2&1\\1&2\end{bmatrix}.
\end{align*}
Then two matrices
\begin{align*}
     \Theta ZBZ^\top \Theta=\begin{bmatrix}0.2&0.1&0.1\\0.1&0.2&0.2\\0.1&0.2&0.2\end{bmatrix}\text{ and } \tilde{\Theta} \tilde{Z}\tilde{B}\tilde{Z}^\top \tilde{\Theta} =\begin{bmatrix}0.05&0.1&0.1\\0.1&0.2&0.2\\0.1&0.2&0.8\end{bmatrix}
\end{align*}
share the same off-diagonal components. 
\end{example}
In this case, even if we observe infinitely many copies of \(A\) and find \(\mathbb{E}(A)\), there is no way to decide whether the model is generated by \(\{Z, \Theta, B\}\) or \(\{\tilde{Z}, \tilde{\Theta}, \tilde{B}\}\), which have different community structures. 

\begin{example}
Consider two parameter sets:
\begin{align*}
\Theta=\begin{bmatrix}1&0&0&0\\0&1&0&0\\0&0&1&0\\0&0&0&1\end{bmatrix},\qquad Z&=\begin{bmatrix}1&0\\1&0\\0&1\\0&1\end{bmatrix},\qquad B=\frac{1}{10}\begin{bmatrix}1&0\\0&4\end{bmatrix};\\
\tilde{\Theta}=\begin{bmatrix}1&0&0&0\\0&1&0&0\\0&0&1&0\\0&0&0&2\end{bmatrix},\qquad \tilde{Z}&=\begin{bmatrix}1&0\\1&0\\0&1\\0&1\end{bmatrix},\qquad \tilde{B}=\frac{1}{10}\begin{bmatrix}1&0\\0&2\end{bmatrix}.
\end{align*}
Then the two matrices
\begin{align*}
     \Theta ZBZ^\top \Theta=\begin{bmatrix}0.1&0.1&0&0\\0.1&0.1&0&0\\0&0&0.4&0.4\\0&0&0.4&0.4\end{bmatrix}\text{ and } \tilde{\Theta} \tilde{Z}\tilde{B}\tilde{Z}^\top \tilde{\Theta} =\begin{bmatrix}0.1&0.1&0&0\\0.1&0.1&0&0\\0&0&0.2&0.4\\0&0&0.4&0.8\end{bmatrix}
\end{align*}
share the same off-diagonal components.  
\end{example}
In this example, although the community structure is well-defined up to a permutation, the degree parameter is not identifiable. 

\subsection{Identifiability of the DCSBM}
We show our main result that the identifiability issue of the DCSBM can be resolved under a mild condition on the size of the smallest community.
\begin{condition}\label{con}
For a DCSBM parametrized by \(\{Z, \Theta, B\}\), assume that each community has at least three members. That is, $Z^{\top}1_n\geq3\cdot1_K$, where the inequality holds entry-wise. 
\end{condition}

\begin{theorem}\label{thm1}
Assume that both parameter systems \(\{Z, \Theta, B\}\) and \(\{\tilde{Z}, \tilde{\Theta},  \tilde{B}\}\) satisfy Condition \ref{con}. These parameters define the same model, i.e., 
\eqref{formula5} holds, if and only if \(\tilde{Z}=ZP\), \(\tilde{B}=P^{\top}DBDP\), and \(\tilde{\Theta}=\Theta \diag\left(ZD^{-1}1_K\right)\) where \(P\) is a \(K\times K\) permutation matrix and \(D\) is a \(K\times K\) positive diagonal matrix.
\end{theorem}

\noindent \textbf{Proof of Theorem \ref{thm1}:} The ``if" part of the statement follows directly from Proposition  \ref{prop1}.
 
For the ``only if" portion, it suffices to show that equation \eqref{formula5} and Condition \ref{con} imply that \(\Theta ZBZ^{\top}\Theta=\tilde{\Theta}\tilde{Z}\tilde{B}\tilde{Z}^{\top}\tilde{\Theta}.\) Then the conclusion follows Proposition \ref{prop1}. 

Note that equation \eqref{formula5} implies that there exists a diagonal matrix \(C\) such that
\[\Theta ZBZ^{\top}\Theta =\tilde{\Theta}\tilde{Z}\tilde{B}\tilde{Z}^{\top}\tilde{\Theta}+C\nonumber,\]
which leads to
\begin{equation}\label{formula6}
\Theta^{-1}C\Theta^{-1} = ZBZ^{\top}-\Theta^{-1}\tilde{\Theta}\tilde{Z}\tilde{B}\tilde{Z}^{\top}\tilde{\Theta}\Theta^{-1}.
\end{equation} 
Let \( e_i \) denote the \( i \)-th coordinate vector in \(\mathbb{R}^n\), defined as the vector with a 1 in the \( i \)-th position and 0 elsewhere. Let $i$ and $j$ be two nodes with $z_i=z_j$. Multiplying  \(e_i-e_j\) on both sides of (\ref{formula6}), we have  
\begin{align*}
\Theta^{-1}C\Theta^{-1}(e_i-e_j) &= \left(ZBZ^{\top}-\Theta^{-1}\tilde{\Theta}\tilde{Z}\tilde{B}\tilde{Z}^{\top}\tilde{\Theta}\Theta^{-1}\right)(e_i-e_j)\\
&=-\Theta^{-1}\tilde{\Theta}\tilde{Z}\tilde{B}\tilde{Z}^{\top}\tilde{\Theta}\Theta^{-1}(e_i-e_j),
\end{align*}
because $Z^{\top}(e_i-e_j)=0$. The last equation further implies
\begin{equation}\label{formula7}
\tilde{\Theta}^{-1}C\Theta^{-1} (e_i-e_j) = -\tilde{Z}\tilde{B}\tilde{Z}^{\top}\tilde{\Theta}\Theta^{-1}(e_i-e_j)=\tilde Zw,
\end{equation}
where \(w=-\tilde{B}\tilde{Z}^{\top}\tilde{\Theta}\Theta^{-1}(e_i-e_j)\).

If \(w\) has a nonzero component, say \(w_l\neq0\), then \(u=\tilde{Z}w\) has at least 3 nonzero components by Condition 1 since $u_s=w_l$ for all nodes $s$ with $\tilde z_s=l$. In contrast, the left-hand side of equation \eqref{formula7} has at most \(2\) nonzero components. Thus, equation \eqref{formula7} indicates \(w=0\), and hence \(C_{ii}=C_{jj}=0\). We can repeat the argument with arbitrary $i$ and $j$ from the same community and conclude $C=0$. \hfill $\Box$

In short, we conclude that for a parameter system $\{Z,\Theta,B\}$ described in Section 2, Condition 1 is necessary and sufficient for the identifiability of the DCSBM.

\subsection{Additional remarks}
In the previous two subsections, we show that the requirement on the minimal community size is crucial for the identifiability of the DCSBM. Here we further clarify how the community size affects the identifiability of the community structure.

\begin{condition}\label{con2}
For a DCSBM parametrized by \(\{Z, \Theta, B\}\), assume that each community has at least two members. That is, $Z^{\top}1_n\geq2\cdot1_K$, where the inequality holds entry-wise. 
\end{condition}

\begin{proposition}\label{prop2}
If both parameter systems \(\{Z, \Theta, B\}\) and \(\{\tilde{Z}, \tilde{\Theta},  \tilde{B}\}\) satisfy Condition \ref{con2}, and  
\eqref{formula5} holds, then \(\tilde{Z}=ZP\). 
\end{proposition}
\noindent \textbf{Proof of Proposition \ref{prop2}:} We show this result by contradiction. If \(\tilde{Z}=ZP\) does not hold, then there are two nodes, say $i$ and $j$, such that $i$ and $j$ belong to the same community in one parameter system, but belong to different communities in the other parameter system. Without loss of generality, assume that two nodes $i,\,j,$ and their community labels satisfy
\begin{itemize}
    \item $ z_i=1$ and $ z_j=2$;
    \item  $\tilde z_i= \tilde z_j=1$. 
\end{itemize}

From the remaining nodes, pick representatives from each group and assign the node and group labels so that node $k$ is in group $k$, i.e., set $z_k=k$ for $k\in \{1,\ldots K\}$ such that node $1$ and node $2$ are distinct from $i$ and $j$, respectively. Condition \ref{con2} guarantees that there are at least two nodes in a group. A figure of the group structure is illustrated below.  

\begin{center}
\begin{tikzpicture}
    \node at (-3,0) {Model 1 $\{Z, \Theta, B\}$};
    \node at (-3,-1.5) {Model 2 $\{\tilde Z, \tilde\Theta,\tilde B\}$};
    
    \draw (0,0) ellipse (0.75 and 0.5) node {$1, i,\cdots$};
    \draw (1.6,0) ellipse (0.75 and 0.5) node {$2, j,\cdots$};
    \draw (3.2,0) ellipse (0.75 and 0.5) node {$3,\cdots$};
    \node at (4.5,0) {$\cdots$};
    \draw (6,0) ellipse (0.75 and 0.5) node {$K,\cdots$};
    
    \node[scale=0.8] at (0,0.75) {Gp. 1};
    \node[scale=0.8] at (1.5,0.75) {Gp. 2};
    \node[scale=0.8] at (3,0.75) {Gp. 3};
    \node[scale=0.8] at (6,0.75) {Gp. $K$};
    
    \draw (0,-1.5) ellipse (0.75 and 0.5) node {$i,j,\cdots$};
    \draw (1.6,-1.5) ellipse (0.75 and 0.5);
    \draw (3.2,-1.5) ellipse (0.75 and 0.5);
    \node at (4.5,-1.5) {$\cdots$};
    \draw (6,-1.5) ellipse (0.75 and 0.5);
\end{tikzpicture}
\end{center}

Note that equation \eqref{formula5} implies \eqref{formula6}, whose right-hand side has zero off-diagonal entries. This gives rise to $n(n-1)/2$ equations of the form 
\begin{equation}\label{nodeeq}
B_{z_pz_q}=\frac{\tilde{\theta}_p\tilde{\theta}_q}{\theta_p\theta_q}\tilde{B}_{\tilde{z}_p\tilde{z}_q},\quad 1\leq p<q\leq n.
\end{equation} 

Applying equation (8) to nodes $i$, $j$, and $1,\dots,\, K$, we have 
\begin{equation*} 
B_{1k}=B_{z_iz_{k}}=\frac{\tilde{\theta}_i\tilde{\theta}_{k}}{\theta_i\theta_{k}}\tilde B_{\tilde{z}_i\tilde z_{k}}
\qquad\text{and}\qquad B_{2k}=B_{z_jz_{k}}=\frac{\tilde{\theta}_j\tilde{\theta}_{k}}{\theta_j\theta_{k}}\tilde B_{\tilde{z}_j\tilde{z}_{k}}.
\end{equation*}
Since $\tilde{z}_i=\tilde{z}_j$, these equations imply that for all $k\in\{1,...,\,K\}$, \[\frac{\tilde{\theta}_j\theta_i}{\tilde{\theta}_i\theta_j}B_{1k}=B_{2k}.\] 
It implies that $B$ is not a full rank matrix, which contradicts the model assumption. \hfill $\Box$
 


Proposition \ref{prop2} indicates that the minimal size requirement for the identifiability of the community structure is two. 
As a corollary of Proposition \ref{prop2}, Condition \ref{con2} also implies the identifiability of the standard SBM. This minimal size requirement cannot be relaxed because of the following example. 

\begin{example}
For a standard SBM parametrized by \(\{Z, B\}\), the expected adjacency matrix $\Delta=ZBZ^\top$. Consider two parameter sets: 

\begin{align*}
Z&=\begin{bmatrix}1&0\\0&1\\0&1\end{bmatrix},\qquad B=\frac{1}{10}\begin{bmatrix}1&0\\0&1\end{bmatrix};\\
\tilde{Z}&=\begin{bmatrix}1&0\\0&1\\0&1\end{bmatrix},\qquad \tilde{B}=\frac{1}{10}\begin{bmatrix}2&0\\0&1\end{bmatrix}.
\end{align*}
Then two matrices
\begin{align*}
     ZBZ^\top=\begin{bmatrix}0.1&0&0\\0&0.1&0.1\\0&0.1&0.1\end{bmatrix}\text{ and } \tilde{Z}\tilde{B}\tilde{Z}^\top =\begin{bmatrix}0.2&0&0\\0&0.1&0.1\\0&0.1&0.1\end{bmatrix}
\end{align*}
are the same up to a diagonal, but $B\neq \tilde{B} $.
\end{example}

\section{Conclusion}
We have clarified identifiability issues on the DCSBM. Our result implies that the community structure of the DCSBM is well-defined under a mild condition on the size of the smallest community. It is worth mentioning that all the theoretical results discussed in this paper rely solely on the mean structure of the model. Therefore, these results are applicable not only to the DCSBM with a Bernoulli distribution on links, but also to networks with a Poisson distribution on links, or even networks with continuous link weights.

\section*{Acknowledgment}
\label{sec:acknowledgement}
The authors thank the Associate Editor for the valuable comments.
This work is partially supported by the National Science Foundation grants DMS-1937229, DMS-2245380 and DMS-2245381.

\bibliographystyle{apalike}
\bibliography{network.bib} 

\end{document}